\documentclass[10pt,journal]{IEEEtran}
\usepackage{epsfig,graphics,graphicx,epsf,color,amsmath,balance,cite,enumerate}

\begin{document}

\title{A CRC-aided Hybrid Decoding for Turbo Codes}

\author{
\authorblockN{Yuejun~Wei, Ming~Jiang, Wen~Chen, and Yuhang~Yang}

\thanks{
Y. Wei is with the Department of Wireless Research, Huawei Technologies, Shanghai, China (email: weiyuejun@huawei.com);

M. Jiang is with National Mobile Communications Research Laboratory, Southeast University, Nanjing, China (e-mail: jiang\_ming@seu.edu.cn);

W. Chen and Y. Yang are with the Department of Electronic Engineering, Shanghai Jiao Tong University, Shanghai, China (e-mail: \{wenchen;yhyang\}@sjtu.edu.cn).}
}
\maketitle

\begin{abstract}
Turbo codes and CRC codes are usually decoded separately according to the serially concatenated inner codes and outer codes respectively. In this letter, we propose a hybrid decoding algorithm of turbo-CRC codes, where the outer codes, CRC codes, are not used for error detection but as an assistance to improve the error correction performance. Two independent iterative decoding and reliability based decoding are carried out in a hybrid schedule, which can efficiently decode the two different codes as an entire codeword. By introducing an efficient error detecting method based on normalized Euclidean distance without CRC check, significant gain can be obtained by using the hybrid decoding method without loss of the error detection ability.
\end{abstract}


\section{Introduction}
Channel codes with iterative decoding methods, such as turbo codes and LDPC codes, are extensively utilized in wireless communication systems. Turbo codes \cite{3GPP2011}, as in the long-term evolution (LTE) protocols, can not only achieve high throughput with its parallel architecture, but also support almost any code rate and arbitrary block length from 40 bits to 6144 bits to cover various services in LTE systems. For the voice over internet protocol (VoIP) service in LTE, the transmission latency and quality of turbo coded data are particular critical. However, turbo codes with short information length usually suffer from severe performance degradation and can not guarantee the quality of the VoIP service, where the information length are limited from 40 bits to about 328 bits.

On the other hand, the cyclic redundancy check (CRC) codes, conventionally for error detection, have no further ability to correct the transmission errors. But they can improve the power consumption and throughput of turbo decoder by early termination, and give indication for retransmission \cite{Shibutani1999VTC-Fall,Chen2005WCNC,Ma2005TWC,Chen2008VTC-Fall}. There are always 24 CRC bits attached after the information bits in LTE physical layer, where the mixed bit stream to be encoded is called a code block (CB). The CRC coding has considerable redundancy which can not be ignored, when the length of a CB is sufficiently short for some specific services like VoIP, especially when the user equipment (UE) is located at the edge of the cell and only small size of CBs can be scheduled due to limited signal to noise ratio \cite{TS36213}. As we can see, due to the CRC redundancy for short length of CBs, there exists an inherent performance gap between turbo decoding and maximum likelihood decoding (MLD) for concatenated turbo-CRC codes.

Besides error detection, CRC codes can also give some help in decoding. Soft list Viterbi algorithm (SLVA) aided convolutional decoding \cite{Seshadri1994TCOM} or turbo decoding \cite{Narayanan1998TCOM} introduces an efficient way to take the advantage of CRC error detection to narrow the performance gap between conventional decoding and MLD. However, the main contribution of SLVA in turbo decoding is limited to the improvement on the error floor. CRC codes can also be decoded by iterative decoding with soft output \cite{Wang2008TCOM}, which are considered as one component code serially concatenated with convolutional codes. Unfortunately, the parity check matrices (PCM) of the classic CRC codes are not appropriate for iterative decoding, since the density of PCM is not sparse enough and four-cycle-free can not be guaranteed at all.

In this letter, we propose a CRC-aided hybrid approach for turbo-CRC codes, where the decoding scheme is hybrid with iterative-based standard turbo decoding (STD) and ordered statistics decoding (OSD) \cite{Fossorier1995TIT,Isaka2004TCOM}. Moreover, the hybrid decoding of concatenated turbo-CRC codes incorporates the CRC bits into the OSD process to further lower the error rate. Since the CRC bits participate in the error correction process, which makes them lose
the error detection ability, we introduce an alternative method for error detection based on the normalized Euclidean distance (NED). Simulation results show that the proposed CRC-aided hybrid scheme can significantly improve the decoding performance of turbo codes with short information length.


\section{The Structure of Turbo-CRC codes}
In LTE turbo code, the two component codes are both recursive systematic convolutional (RSC) codes with the same generator polynomial,
\begin{equation}
\label{Gen_Poly_RSC}
{\bf{g}}^{RSC}\left( D \right) = \left[ {1,\frac{{1 + D + {D^3}}}{{1 + {D^2} + {D^3}}}} \right].
\end{equation}
Furthermore, the generator matrix of LTE turbo codes can be written as follows,
\begin{equation}
\label{Gen_Mat_Turbo}
{{\bf{G}}^{\textit{turbo}}} = \left[ {{{\bf{I}}_{k}}\;|\;{{\bf{P}}_{k}}\;|\;{{\tilde {\bf{P}}}_{k}}} \right],
\end{equation}
where the parameter $k$ is the CB length. The turbo encoded bits are associated with identity matrix ${\bf{I}}_{k}$ and matrix ${\bf{P}}_{k}$ (the interleaved one as ${{\tilde {\bf{P}}}_{k}}$), which are corresponding to the systematic bits and parity bits, respectively.

According to the second entry in \eqref{Gen_Poly_RSC}, we can transform the polynomial fraction into an infinite periodic polynomial,
\begin{equation}
\label{Gen_Poly_RSC_Periodic}
\begin{aligned}
\frac{1 + D + D^3}{1 + D^2 + D^3} &= 1 + \sum\limits_{i=0}^\infty {\sum\limits_{j=1}^7 {a_jD^{7i+j}}}.
\end{aligned}
\end{equation}
Then, an infinite binary sequence ${\bf{A}}$ can be obtained from the coefficients in \eqref{Gen_Poly_RSC_Periodic},
\begin{equation}
\label{Gen_Poly_Sequence}
{\bf{A}}=\{{1, \bf{a}}, {\bf{a}}, {\bf{a}}, ...... \},
\end{equation}
where the vector ${\bf{a}}=[a_1,a_2,..., a_7]=[1110010]$.

Then, we can get a column vector ${\bf{p}}_0$ by selecting the first $k$ elements from ${\bf{A}}$ in line with the length of CB. From the column vector ${\bf{p}}_0$ with dimension $k$, the other $k-1$ column vectors ${\bf{p}}_1, ..., {\bf{p}}_{k-1}$ can also be obtained after a sequential process step by step,
\begin{equation}
\label{Gen_matrix_P_Col}
{\bf{p}}_{i+1} = [{\bf{\Theta}}, {\bf{e}}_0, {\bf{e}}_1, {\bf{e}}_2, ..., {\bf{e}}_{k-2}]^T {\bf{p}}_{i}, i=0, ..., k-2,
\end{equation}
where $[\cdot]^T$ denotes the transpose operation and ${\bf{\Theta}}$ is an all-zero column vector with dimension $k$. The other $k-1$ column vectors with dimension $k$ in \eqref{Gen_matrix_P_Col}, ${\bf{e}}_i, i=0, ..., k-2$, are the indicator vectors with one in the $i$-th position and zeros elsewhere.

The sub-matrix ${\bf{P}}_{k}$ in generator matrix ${\bf{G}}_{turbo}$ can be built from the $k$ column vectors ${\bf{p}}_0, ..., {\bf{p}}_{k-1}$,
\begin{equation}
\label{Sub_Matrix_P}
{\bf{P}}_{k} = [{\bf{p}}_0, {\bf{p}}_{1}, {\bf{p}}_{2},    ..., {\bf{p}}_{k-1}]^T.
\end{equation}
The other sub-matrix ${\tilde {\bf{P}}}_{k}$ can be obtained by permuting the rows of the matrix ${\bf{P}}_{k}$ according to the QPP interleaver \cite{3GPP2011}.

Besides turbo coding, the CRC process is also a mandatory requirement in LTE. Two different CRC schemes with the same length of parity bits (24 bits) are specified in the protocol for each CB and each transport block, respectively. When the length of the information bits is less than 6120 bits, only one 24-bit CRC encoding, named as CRC24a in \cite{3GPP2011}, is carried out followed by the turbo encoding.


Considering the 24 CRC bits before turbo encoding, the real length of the information bits is $m=k-24$ and the entire generator matrix of the turbo-CRC code is the product of the two generator matrices of the turbo code and CRC code,
\begin{equation} \label{Gen_Mat_Turbo_CRC}
{\bf{G}} = {\bf{G}}^{CRC} \times {\bf{G}}^{turbo},
\end{equation}
where ${\bf{G}}^{CRC}$ should take the systematic format of the generator matrix.
However, we usually can only obtain the non-systematic generator matrix ${\hat {\bf{G}}}^{CRC}$ by the generator polynomial of the CRC code in LTE,
\begin{equation}
{\bf{g}}^{CRC} \left( {D} \right) = \sum\limits_{i=0}^{24}{g_i \times D^i},
\end{equation}
where $\{g_{24}, g_{23}, ..., g_0\}=\{1100001100100110011111011\}$, and the matrix ${\hat {\bf{G}}}^{CRC}$ can be represented by the coefficients of the ${\bf{g}}^{CRC} \left( {D} \right)$ as follows,
\begin{equation}\label{Gen_Mat_CRC}
\begin{small}
{\hat{\bf{G}}}^{CRC} = {\left[ {\begin{array}{*{20}{c}}
{{g_0}}&{{g_1}}& \cdots &{{g_{24}}}&{}&{}&{}\\
{}&{{g_0}}&{{g_1}}& \cdots &{{g_{24}}}&{}&{}\\
{}&{}& \ddots & \ddots & \ddots & \ddots &{}\\
{}&{}&{}&{{g_0}}&{{g_1}}& \cdots &{{g_{24}}}
\end{array}} \right]_{m \times k}.}
\end{small}
\end{equation}

Then, the Gaussian elimination is carried out on the matrix ${\hat {\bf{G}}}^{CRC}$ to get the systematic formation,
\begin{equation}\label{Systematic_Gen_Mat_CRC}
{\bf{G}}^{CRC} = \lbrack {{{\bf{I}}_{m \times m}}|{{\bf{Q}}_{m \times 24}}} \rbrack.
\end{equation}

\section{CRC-aided Hybrid Decoding Algorithm}

\subsection{The Hybrid Decoding with STD and OSD}

The reliability-based OSD and the iterative-based STD in hybrid decoding can be carried out individually and collaboratively. These two decoding processes are in terms of the different principles in error-correcting. Hence, the STD and the OSD both can be carried out independently with the signals received from the channel, $Y=\{y_1, y_2, \cdots, y_{3k+12}\}$.

The OSD process can be divided into three major parts, sorting, Gaussian elimination and re-encoding. Firstly, the hard-decision bits based on the signs of the input information $\{R_1, \cdots, R_{3k}\}$ and the columns of the $G^{turbo}$ are both permuted according to the reliabilities $\{|R_1|, \cdots, |R_{3k}|\}$ sorted in decreasing order. Then, the Gaussian elimination is applied to the permutated matrix to obtain the systematic generator matrix $\tilde G^{turbo}$. Finally, the re-encoding are executed according to the $\tilde G^{turbo}$ and the permutated hard-decision bits, where the CRC detection is performed based on the first reverse-permutated $k$-bits in the re-encoded codeword.

 In this paper, the Max-Log-MAP algorithm is utilized for STD due to its relatively low complexity and good performance. After each iteration, not only the log-likelihood-ratios (LLRs) of the systematic bits, but also those of two sets of the parity bits should be calculated for the hybrid decoding. The updated LLRs of both the systematic bits and the parity bits after the $t$-th iteration are denoted as $\{L_1^t, \cdots, L^t_{3k}\}$, which can be expressed as follows,
\begin{align}\label{eq_LLR_sys}
\begin{array}{r}
L_i^t = {{\mathop {{\rm{max}}}\limits_{m,{m'}} }}\left( {{\gamma_i^{{m'},m}}+ {\alpha_{i-1}^{m'}} + {\beta_i^m}} \right)|({d_i} = 0)\\
               - {{\mathop {{\rm{max}}}\limits_{m,{m'}} }}\left( {{\gamma_i^{{m'},m}}+ {\alpha_{i-1}^{m'}} + {\beta_i^m}} \right)|({d_i} = 1),
\end{array}
\end{align}
where $d_i$ denotes the systematic bit or parity bit, and $\alpha$, $\beta$ and $\gamma$ denote the forward recursive,
backward recursive and branch transition probabilities,
respectively.

 The updated LLRs after each iteration of STD can provide more reliable information, so the OSD can be efficiently scheduled using the output LLRs of the $t$-th iteration in STD, $\{R_i = L_i^t, i=1, \cdots, 3k\}$. A parameter $f$ is selected to indicate the iteration in which the OSD process is launched and executed in subsequent each and every iteration. For example, when the parameter $f$ is set to $1$ and maximum number of iterations $T$ respectively, the OSD process is performed after each iteration and only once after the last iteration of the turbo decoding.

As the LLR oscillation phenomenon is an inevitable problem during the iterative turbo decoding \cite{Gounai2005IEICE-FUND}, the magnitudes of the updated LLRs in current iteration are not always the reliable metric for the OSD process. In order to overcome the LLRs oscillations, the accumulation of LLRs is proposed for the OSD process of LDPC codes \cite{Jiang2007ComLett}. After $t, 1 \le t \le T$ iterations, the accumulated LLRs are set to be the input information of the OSD, which can be calculated as follows,
\begin{equation}\label{LLR_Accumulation}
R_i^t = L_i^t + \alpha R_i^{t-1}, \,\, i=1, \dots, 3k,
\end{equation}
where the initialized accumulated LLRs, $\{R_1^0, \cdots, R_{3k}^0\}$, are all set to zero. The accumulated LLRs are equivalent to the updated LLRs in each iteration when the accumulation parameter $\alpha$ is set to zero.

\subsection{The CRC Assistance in OSD}

For the short turbo codes in LTE, the $24$-bits redundancy caused by CRC codes can not be ignored. Although the CRC codes are usually used for error detection, they can still be treated as component codes of the concatenated turbo-CRC codes and make extra contributions to the error correction. We replace the matrix $G^{turbo}$ with the entire generator matrix of turbo-CRC codes ${\bf{G}}$ in \eqref{Gen_Mat_Turbo_CRC} in OSD process, incorporating the additional constraints introduced by the CRC codes in decoding. With the assistance of the CRC codes, the hybrid decoding can further narrow the performance gap to the MLD of turbo-CRC codes with $k-24$ information bits, when the Gaussian elimination and re-encoding in the OSD process is carried out according to the matrix ${\bf{G}}$.

All the re-encoded codewords from the OSD process according to the matrix ${\bf{G}}$ always satisfy the CRC check, since the matrix ${\bf{G}}$ already incorporates the CRC generator matrix. Then, we can only select the codeword with the minimum Euclidean distance to the received signals as the optimal output. Assuming that order-$N$ re-encoding is performed and there are total $\Phi(N)$ codewords generated, the computation of Euclidean distance for the $n$-th codeword $C^n = \{c^n_1, \cdots, c^n_{3k}\}$ can be written as follows,
\begin{equation}\label{Euclidean-distance}
d^n = \sum_{c^n_i \ne z_i, i=1, \cdots, 3k}{|y_i|}, n=1, \cdots, \Phi(N),
\end{equation}
where the sequence $\{z_1, z_2, \cdots, z_{3k}\}$ is the hard decision of the received signal $Y$. The codeword $C^*$ which has the minimum distance
\begin{equation}\label{Min-Euclidean-distance}
d^* = \min \limits_{n=1, \cdots, \Phi(N)}{d^n}
\end{equation}
is regarded as the final result, when the STD fails after $T$ iterations.


\subsection{An Alternative Error Detection Method}

In the CRC-aided hybrid decoding, there is always a valid output from the OSD process for the turbo-CRC codes, since the CRC bits have participated in the decoding process and lost the error detection ability. Therefore, the increase of the undetected error rate (UER) is unavoidable, which will lead to a serious degradation of the system throughput. To make up for the deficiency in error detection, we propose an alternative detection criterion based on the NED. The NED metric of the optimal codeword $C^*$ is calculated as follows,
\begin{equation}\label{NED}
\hat d^* = \frac{d^*}{\sum \limits_{i=1, \cdots, 3k}{|y_i|}}.
\end{equation}

A threshold $\eta$ according to the NED metrics can be set for error detection. If the NED metric of the output codeword $C^*$ from the OSD is larger than the preset threshold, i.e., $\hat d^* >\eta$, the hybrid decoding result is regarded as error. Not only the undetected error but also the false alarm probability should be considered in the optimization of $\eta$. The UER due to the NED criterion with the optimized $\eta$ can still achieve near one thousandth of the total error rate, although the error detection ability of the NED criterion is obviously worse than that of the 24-bit CRC. Furthermore, the normalized calculations make the NED metrics independent of the scaling of the received signals. Simulation results also show that the difference of SNR has little impact on the optimal value of the threshold $\eta$.


\subsection{Complexity Analysis}

The computational complexity of the OSD algorithm is close to $O(k^3)$, because as the kernel process, the Gaussian elimination of the generator matrix has a cubic complexity. Moreover, the Gaussian elimination is not suitable for parallel implementation and can not match the high-speed turbo decoding in LTE system. Fortunately, our proposed hybrid decoding is mainly applied for the turbo codes with short information length. When the information length of turbo codes is within the range of about 352 for VoIP services, the total decoding delay and the computational complexity of the hybrid decoding is acceptable.

After the $t$-th turbo iteration, the OSD process with high-order can also be taken to further improve the error performance, where the re-encoding process is in terms of the different combinations of the $k$ hard decision bits with the highest reliability. However, only order-$0/1/2$ OSD process is usually utilized due to the fast growth in the number of combinations.

\section{Simulation results}

In order to compare the performance of different decoding algorithms, the STD, the hybrid decoding and the CRC-aided hybrid decoding schemes are employed. The STD is using the enhanced Max-Log-MAP algorithm with a scaling factor of 0.75 over the extrinsic information \cite{Gracie:99,Wei2012}. The maximum number of iterations of STD is set to be $T=8$. The OSD($N,f,\alpha$) denotes the order-$N$ OSD process carried out after the $f$-th iteration, where the factor $\alpha$ is set for the LLR accumulation. To show the improvements on the short-length turbo codes, we set the lengths of systematic bits to be $k=40$ and $96$, respectively, where the 24-bits CRC is applied and the lengths of the original information bits are $m=16$ and $72$, respectively. With the additional 12 tail bits, the actual rates of the turbo-CRC codes are $4/33$ and $6/25$ accordingly. The frame error rates (FER) of the different decoding algorithms are simulated under the additive white Gaussian noise (AWGN) channel with binary phase shift keying (BPSK) modulation.

\begin{figure}[!t]
\centerline{\psfig{file=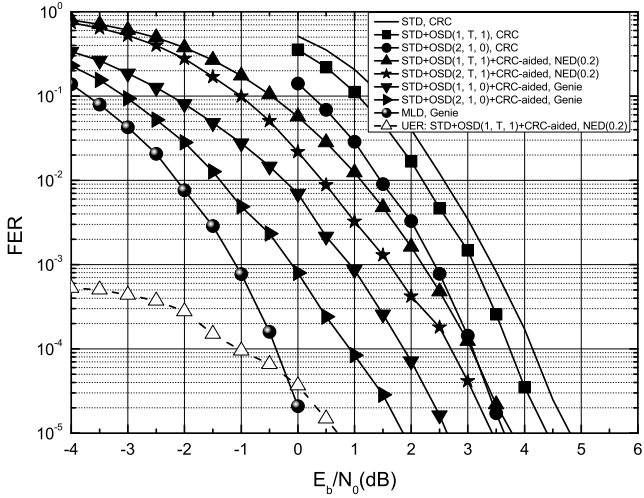,width=4.35in}}
\centering
\caption{Error performance of turbo-CRC code with $k=40$ by using different decoding algorithms and error detection criteria, $\eta=0.2$.}
\label{fig:k40}
\end{figure}

The error performances of the turbo-CRC code with $k=40$ by using different decoding algorithms are shown in Fig. \ref{fig:k40}. The hybrid decoding with CRC as error detection (`STD+OSD(2,1,0)') obtains less than 1dB gain over the STD, in which the order-2 OSD process is carried out after each iteration of the STD. With the order-$2(1)$ OSD approach only carried out after the last iteration, more than 2dB(1.5dB) gain can be obtained by the CRC-aided hybrid decoding, which is much bigger than the hybrid decoding. According to the simulation, by the proposed NED error detection scheme with $\eta=0.2$, the UER (dotted line) is always lower than $10^{-3}$. For services like VoIP, the FER is usually lower than $10^{-2}$, which means the $E_b/N_0$ is higher than 1dB. At this condition, the UER is lower than $10^{-5}$, which is low enough for VoIP services.
In order to further show the decoding gain of the CRC-aided hybrid decoding, we give the performance of the CRC-aided hybrid decoding with genie-aided error detection (`STD+OSD(2, 1, 0)+CRC-aided, Genie'), which can approach the optimal performance of MLD within 0.5dB when FER is $10^{-3}$.

\begin{figure}[!t]
\centerline{\psfig{file=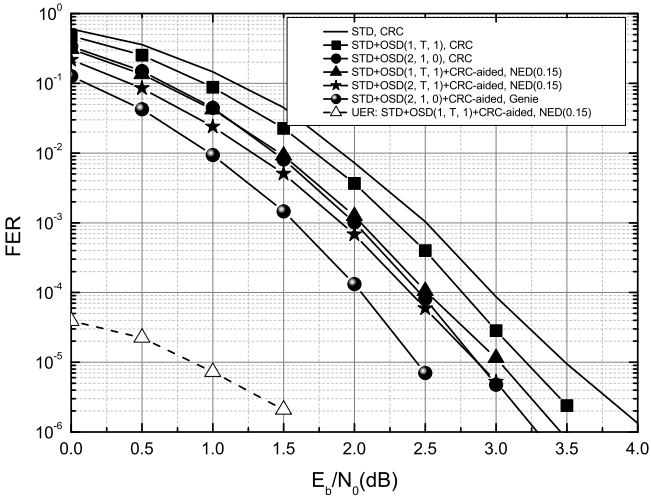,width=4.35in}}
\centering
\caption{Error performance of turbo-CRC code with $k=96$ by using different decoding algorithms and error detection criteria, $\eta=0.15$.}
\label{fig:k96}
\end{figure}

\mbox{Fig. \ref{fig:k96}} shows the error performances of the turbo-CRC code with \mbox{$k=96$} by using different decoding algorithms. Although the redundancy ratio of the CRC codes decreases with the increase of the length of $k$, there is still a noticeable performance gain for the proposed scheme.
The CRC-aided hybrid decoding with order-1 OSD carried out after the last iteration of STD (`STD+OSD(1, T, 1)+CRC aided, NED(0.15)') has almost the same performance with the hybrid decoding with order-2 OSD carried out after each iteration of STD (`STD+OSD(2, 1, 0), CRC'), and the gain over STD is about 0.5dB. The UER (dotted line) caused by NED criterion is much lower than the FER and can be negligible. We note that the CRC-assisted hybrid decoding with genie-aided error detection can only obtain additional 0.25dB gain over that with NED as the error detection.

\section{Conclusion}
In this letter, a CRC-aided hybrid decoding algorithm is proposed, in which the CRC bits are not utilized as error detection but as error correction in the OSD process. An alternative error correction criterion based on the normalized Euclidean distance is also proposed in case that the CRC bits have lost the error detection ability. Simulation results show that our proposed scheme can significantly improve the error performance of the turbo-CRC codes with short information lengths. At the same time, the undetected error rates are pretty low for practice applications such as VoIP services.


\begin{thebibliography}{1}
\bibitem{3GPP2011}
3GPP, ``TS36.212v10.2.0: Multiplexing and channel coding," Jun. 2011.
\bibitem{Shibutani1999VTC-Fall}
A. Shibutani, H. Suda, and F. Adachi, ``Complexity reduction of turbo decoding," {\em IEEE Vehicular Technology Conference}, vol. 3, 1999-Fall, pp. 1570-1574.
\bibitem{Chen2005WCNC}
H. Chen, L. Cao, and C. W. Chen, ``Constrained decoding for turbo-CRC code with high spectral efficient modulation," {\em IEEE Wireles
Communications and Networking Conference}, vol. 2, 2005, pp. 1050-1054.
\bibitem{Ma2005TWC}
Z. Ma, W. H. Mow, and P. Fan, ``On the complexity reduction of turbo decoding for wideband CDMA," {\em IEEE Trans. Wireless Commun.}, vol. 4, no. 2, pp. 353-356, Mar. 2005.
\bibitem{Chen2008VTC-Fall}
J. F. Chen, ``Lwo-level early stopping algorithm for LTE turbo decoding," {\em IEEE Vehicular Technology Conference}, 21-24 Sept. 2008-Fall.
\bibitem{Seshadri1994TCOM}
N. Seshadri and C.-E. W. Sundberg, ``List Viterbi decoding algorithms with applications," {\em IEEE Trans. Commun.}, vol. 42, no. 2/3/4, pp. 313-323, Feb. /Mar. /Apr. 1994.
\bibitem{TS36213}
3GPP, ``TS36.213v10.2.0: Physical layer procedures," Jun. 2011.
\bibitem{Narayanan1998TCOM}
K. R. Narayanan and G. L. Stuber, ``List decoding of turbo codes," {\em IEEE Trans. Commun.}, vol. 46, no. 6, pp. 754-762, Jun. 1998.
\bibitem{Wang2008TCOM}
R. Wang, W. Zhao, and G. Giannakis, ``CRC-assisted error correction in a convolutionally coded system," {\em IEEE Trans. Commun.}, vol. 56, no. 11, pp. 1807-1815, Nov. 1998.
\bibitem{Fossorier1995TIT}
M. P. C. Fossorier and S. Lin, ``Soft-decision decoding of linear block codes based on ordered statistics," {\em IEEE Trans. Inf. Theory}, vol. 41, no. 5, pp. 1379-1396, Sep. 1995.
\bibitem{Isaka2004TCOM}
 M. Isaka, M. Fossorier, and H. Imai, ``On the suboptimality of iterative decoding for turbo-like and LDPC codes with cycles in their graph representation, {\em IEEE Trans. Commun.}, vol. 52, no. 5, pp. 845-854, May 2004.
\bibitem{Gounai2005IEICE-FUND}
S. Gounai and T. Ohtsuki, ``Decoding algorithms based on oscillation for low-density parity check codes," {\em IEICE Trans. Fund.}, vol. E88-A, no. 8, pp. 2216-2226, Aug. 2005.
\bibitem{Jiang2007ComLett}
M. Jiang, C. Zhao, E. Xu, and L. Zhang, ``Reliability-based iterative decoding of LDPC codes using likelihood accumulation," {\em IEEE Commun. Lett.}, vol. 11, no. 8, pp. 677-679, Aug. 2007.
\bibitem{Gracie:99}
K.~Gracie, S.~Crozier, and A.~Hunt, ``Performance of a Low-Complexity Turbo Decoder with a Simple
Early Stopping Criterion Implemented on a SHARC Processor,"
{\em Proceedings of the 6th International Mobile Satellite Conference (IMSC '99)},
Ottawa, Ontario, Canada, pp.~281-286, Jun.~1999.
\bibitem{Wei2012}
Y.~Wei, Y.~Yang, L.~Wei and W.~Chen, ``Comments on `A New Parity-Check Stopping Criterion for Turbo Decoding',"
{\em IEEE Commun. Lett.}, vol.~16, pp.~1664-1667, Oct. 2012.


\end{thebibliography}


\end{document}